\documentclass[prb,aps,amsmath,amssymb,twocolumn,superscriptaddress,10pt,showkeys, author-year]
{revtex4-2}

\usepackage{amsmath}
\usepackage{amssymb}
\usepackage{amsfonts}

\usepackage{graphicx}

\usepackage{hyperref}
\hypersetup{
breaklinks=true,
colorlinks=true,
citecolor=blue,
linkcolor=red,
filecolor=blue,
urlcolor=blue,
}

\usepackage{orcidlink}

\begin{document}

\title{Bayesian Time-Lapse Full Waveform Inversion \\using Hamiltonian Monte Carlo}
\author{P. D. S. de Lima\orcidlink{0000-0002-7353-536X}}
\affiliation{Departamento de Física Teórica e Experimental, \\Universidade Federal do Rio Grande do Norte, 59078-970 Natal-RN, Brazil}
\affiliation{School of Physics, Trinity College Dublin, Dublin 2, Ireland}%

\author{M. S. Ferreira\orcidlink{0000-0002-0856-9811}}
\affiliation{School of Physics, Trinity College Dublin, Dublin 2, Ireland}
\affiliation{Centre for Research on Adaptive Nanostructures and Nanodevices (CRANN) \& Advanced Materials and Bioengineering Research (AMBER) Centre, Trinity College Dublin, Dublin 2, Ireland}

\author{G. Corso\orcidlink{0000-0003-1748-4040}}
\affiliation{Departamento de Biofísica e Farmacologia, \\ Universidade Federal do Rio Grande do Norte, 59078-970 Natal-RN, Brazil}%

\author{J. M. de Araújo\orcidlink{0000-0001-8462-4280}}%
\affiliation{Departamento de Física Teórica e Experimental, \\Universidade Federal do Rio Grande do Norte, 59078-970 Natal-RN, Brazil}%

\date{\today}

\begin{abstract}
Time-lapse images carry out important information about dynamic changes in Earth's interior which can be inferred using different Full Waveform Inversion (FWI) schemes. The estimation process is performed by manipulating more than one seismic dataset, associated with the baseline and monitors surveys. The time-lapse variations can be so minute and localised that quantifying the uncertainties becomes fundamental to assessing the reliability of the results. The Bayesian formulation of the FWI problem naturally provides confidence levels in the solution, but evaluating the uncertainty of time-lapse seismic inversion remains a challenge due to the ill-posedness and high dimensionality of the problem. The Hamiltonian Monte Carlo (HMC) can be used to effectively sample over high dimensional distributions with affordable computational efforts. In this context, we propose a probabilistic Bayesian sequential approach for time-lapse FWI using the HMC method. Our approach relies on the integration of the baseline survey information as prior knowledge in the monitor estimation. We compare the proposed methodology with a parallel scheme in perfect and perturbed acquisition geometry scenarios. We also investigate the correlation effect between baseline and monitor samples in the propagated uncertainties. 
The results show that our strategy provides accurate times-lapse estimates with errors of similar magnitude to the parallel methodology.
\end{abstract}

\maketitle

\section{Introduction}

Extracting information from noisy and limited data is paramount in subsurface characterization procedures.
In particular, the availability of time-lapse data, that is, multiple seismic data acquired at different calendar times, offers not only the opportunity to determine static structures but also dynamic changes in geological models~\citep{lumley2001}. In this context, the first seismic survey is called baseline while the remaining ones are referred to as monitors. Constructing time-lapse images is crucial
to control the oil production behaviour~\citep{WERNECK2022109937} in reservoirs or to monitor the CO2 storage~\citep{dongli2021, nakata2022} in the injection process. 

Regardless of its application, time-lapse models can be constructed using the Full Waveform Inversion (FWI)~\citep{tarantola1984, virieux2009}. This high-resolution technique takes into account both the phase and amplitude of the wavefield in the model parameter estimation. However, additional challenges are present in the time-lapse inversion due to the small and spatially localized nature of variations combined with the non-repeatability of the acquisition geometry and climate factors~\citep {zhou2021}. 

Several deterministic time-lapse FWI methodologies have been proposed in the literature. The parallel strategy~\citep{lumley2003, plessix2010} independently inverts both baseline and monitor data from the same initial model. The double-difference~\citep{watanabe2005, zhang2013, yang2015, yang2016} inverts directly from data difference while the jointly-inversion~\citep{zheng2012, maharramov2014, maharramov2015, maharramov2016} obtains both models simultaneously. Another approach is the sequential difference~\citep{routh2012, asnaashari2012, Asnaashari, lumley2017}, which uses the inverted baseline model as the initial model in the monitor inversion. Generalizations of the sequential scheme have been recently designed to overcome the non-repeatability problem~\citep{lumleycentraldiff, mardan2023}. However, the uncertainty regarding the time-lapse estimates cannot be properly inferred using these deterministic approaches~\citep{sambridge2002}. In fact, the ill-posed nature of the FWI problem produces solutions with nontrivial propagated errors, such as noise in data, model parametrization and simplification of the physical theory~\citep{tarantola2005}. Therefore, the uncertainty quantification process is fundamental for evaluating result reliability and facilitating the decision-making process.

Bayesian approaches address the challenge of uncertainty quantification by transforming the seismic inversion into a statistical inference problem, where the results become probability density functions which are sampled from the model space. Markov Chain Monte Carlo (MCMC) has been used over decades in geophysical inversion problems~\citep{sambridge2002, sambridge2013, sen_stoffa_2013} for this task. However, MCMC methods are inefficient in estimating probability distributions in high-dimensional spaces, due to the well-known curse of dimensionality effect. This dimensionality issue is substantially suppressed when the Hamiltonian Monte Carlo (HMC) method is used~\citep {betancourt2018conceptual}. The HMC is a sampling technique that explores the geometry of target distribution to avoid the random walk behaviour of standard MCMC methods~\citep{neal2011}. Originally formulated for quantum chromodynamics~\citep{DUANE1987216}, the HMC has been popularized in the seismic inversion community in the last decade~\citep{biswas2017, Fichtner2018, fichtner2018_2, Gebraad2020, Aleardi2020, aleardi2020b, DELIMA2023128618}. Alternatively to MCMC-based methods, the ensemble-based~\citep {eikrem2019, thurin2019, huang2020, eidsvik2020} and variational~\citep{zhang2020imaging, zhang2020seismic, curtis2020, curtis2021, zhang3d2023, zhang2023bayesian} approaches have also been employed to address similar problems.

Advances in Bayesian time-lapse estimation have been performed in~\citep{kotsi2020a, kotsi2020b}, where the uncertainties are estimated directly from data changes using data compression techniques~\citep{aleardi2023}. However, this approach is valid only when the acquisition geometry is repeatable, a strong assumption for real-data applications. In Ref. \citep{zhang2023bayesian}, Zhang and Curtis performed a comparison between the separate and jointly Bayesian time-lapse inversion using a stochastic Stein Variational Gradient Descent (sSVGD) method~\citep{zhang3d2023} by imposing a prior distribution directly in time-lapse estimates. In this paper, we explore a different avenue by considering the posterior information regarding the baseline survey to design a prior distribution to be used in the monitor estimation. Results for this sequential strategy in a simplified scenario where the baseline model is known was reported in~\citep{delima_eage_2023}. Here, we follow a general approach and demonstrate the benefits of employing informative prior information in time-lapse studies by comparing the sequential and parallel approaches in terms of estimates, propagated uncertainties and sample correlation. Moreover, we also consider in our investigations the nonrepeatability effects associated with different source locations in the surveys.

In the following section, we describe the Bayesian formulation of the FWI problem and extensions to the time-lapse case considering the parallel and sequential strategies. We also present the HMC as a sampling method and details involving the tuning strategy to the mass matrix in this section. In section 4, we compare the statistical estimates obtained using both parallel and sequential strategies. We also compare the Bayesian inversion results with deterministic ones. We conclude by discussing the limitations of sequential strategy and possible improvements.

\section{Methods}
\subsection{Bayesian Full Waveform Inversion}

The solution of a FWI problem from a Bayesian perspective is not a single model but a probability density function of models~\citep{tarantola2005}, called the posterior distribution $\rho(\mathbf{m}|\mathbf{d})$. The posterior distribution combines the prior knowledge of the model $\mathbf{m}$ with the information acquired through data measurements $\mathbf{d}^{\text{obs}}$ via Bayes's theorem~\citep{bayes1763}: 
\begin{equation}
    \rho(\mathbf{m}|\mathbf{d}) = \frac{\rho(\mathbf{m})\rho(\mathbf{d}|\mathbf{m})}{\rho(\mathbf{d})}\,, \label{post_dist}
\end{equation}
where $\rho(\mathbf{m})$ is the prior probability distribution which encapsulates the previous model knowledge and the likelihood function $\rho(\mathbf{d}|\mathbf{m})$ that encompasses the conditional probability of an observation being explained by a certain model. The denominator in equation \eqref{post_dist} is called the Bayesian evidence (or partition function): 
\begin{equation}
    \rho(\mathbf{d}) = \int_{\mathbb{M}}\rho(\mathbf{d}|\mathbf{m})\rho(\mathbf{m})\,d\mathbf{m}\,,
\end{equation}
which normalizes the posterior distribution.

In this regard, the set of plausible model parameters $\mathbf{m}$ spans a model space $\mathbb{M}$ and the correspondent mapping to the data space $\mathbb{D}$ is done by a nonlinear forward operator $\mathbf{d}^{\text{mod}}(\mathbf{m}) = \mathbf{G}(\mathbf{m})$. The error between observations and data predictions from a proposal model is called residuals $\Delta\mathbf{d} = \mathbf{d}^{\text{mod}} - \mathbf{d}^{\text{obs}}$. In light of the central limit theorem, we assume that the residual distribution is Gaussian. Therefore, the likelihood function can be written as:
\begin{equation}
    \rho(\mathbf{d}|\mathbf{m}) = \exp{\left[-\frac{1}{2}\Delta\mathbf{d}^{\text{T}}(\Sigma^{\text{noise}})^{-1}\Delta\mathbf{d}\right]}\,, \label{likelihood}
\end{equation}
where $\Sigma^{\text{noise}}$ is the noise covariance matrix. In our synthetic studies, we assume that the data residual error variance is known, even though the residual error distribution in more realistic cases can be estimated~\citep{bodin2012} or may take different forms~\citep{brossier2010, metivier2016, liu2016, carvalho2021}. We notice that in the uncorrelated data case, the matrix $\Sigma^{\text{noise}}$ has only diagonal elements. The statistical analysis of the posterior distribution is typically evaluated by computing the posterior mean
\begin{equation}
    \mathbf{m}^{\text{mean}} = \int_{\mathbb{M}}\mathbf{m} \> \rho(\mathbf{m}|\mathbf{d})\,d\mathbf{m}\,,
\end{equation}
and the posterior covariance
\begin{equation}
   \Sigma_{\textbf{m}} = \int_{\mathbb{M}}(\mathbf{m} - \mathbf{m}^{\text{mean}})^{\text{T}}(\mathbf{m}-\mathbf{m}^{\text{mean}})\rho(\mathbf{m}|\mathbf{d})\,d\mathbf{m}
\end{equation}
of the model parameter $\mathbf{m}$. However, in practical FWI problems we have only $n$ representative model samples $(\mathbf{m}^{(1)}, \mathbf{m}^{(2)} \ldots, \mathbf{m}^{(n)})$ as estimation of the posterior distribution $\rho(\mathbf{m}|\mathbf{d})$. 

\subsection{Bayesian Time-Lapse Full Waveform Inversion}

As far as time-lapse problems are concerned, we need to deal with two different data spaces, each one associated with different seismic acquisitions, namely the baseline $\mathbf{d}^{\text{obs}}_{\text{B}}$ and the monitor $\mathbf{d}^{\text{obs}}_{\text{M}}$ data. Deterministically, this information can be used to construct the baseline $\mathbf{m}_{\text{B}}$ and monitor $\mathbf{m}_{\text{M}}$ models, although we are primarily interested in the time-lapse $\mathbf{m}_{\text{TL}} = \mathbf{m}_{\text{M}} - \mathbf{m}_{\text{B}}$ model change. Accordingly, in the probabilistic realm we need to handle two probability distributions to infer time-lapse changes, the baseline $\rho(\mathbf{m}_{\text{B}}|\mathbf{d}_{\text{B}}) \equiv \rho_{\text{B}}(\mathbf{m}|\mathbf{d})$ and the monitor $\rho(\mathbf{m}_{\text{M}}|\mathbf{d}_{\text{M}}) \equiv \rho_{\text{M}}(\mathbf{m}|\mathbf{d})$ posterior distributions.

The straightforward Bayesian inference approach is to estimate the uncertainty on the time-lapse changes $\mathbf{m}_{\text{TL}}$ directly from data difference $\mathbf{d}_{\text{TL}} = \mathbf{d}_{\text{M}} - \mathbf{d}_{\text{B}}$~\citep{ouair2006}. Although this strategy avoids separately sampling each distribution, thus reducing the computational cost, it cannot be applied in realistic scenarios with nonrepeatable acquisitions. 

Looking at the two distributions separately, one possible strategy is to estimate the baseline $\rho_{\text{B}}(\mathbf{m}|\mathbf{d})$ and monitor $\rho_{\text{M}}(\mathbf{m}|\mathbf{d})$ using the same prior information, which means $\rho_{\text{B}}(\mathbf{m}) = \rho_{\text{M}}(\mathbf{m}) = \rho(\mathbf{m})$. Alternatively, one can follow a joint Bayesian inversion~\citep{zhang2023bayesian}, where baseline and time-lapse probabilities are obtained simultaneously. However, as the time-lapse change is small and spatially localized and the time interval between the acquisition is long enough to construct well-resolved velocity models, it is natural to use all information learned from the baseline inversion as priors for the monitor estimation. Therefore, we follow a probabilistic version of the sequential strategy, where the prior probability used in the monitor inversion is derived from the baseline posterior distribution.

Following this sequential scheme, we first draw $n_{\text{B}}$ samples $(\mathbf{m}^{(1)}_{\text{B}}, \mathbf{m}^{(2)}_{\text{B}}, \ldots, \mathbf{m}^{(n_{\text{B}})}_{\text{B}})$ of the baseline posterior $\rho_{\text{B}}(\mathbf{m}|\mathbf{d})$ which are then used to design a prior distribution to the monitor, that is, we set $\rho_{\text{M}}(\mathbf{m}) = \rho_{\text{B}}(\mathbf{m}|\mathbf{d})$. Using this prior, we collect $n_{\text{M}}$ samples $(\mathbf{m}^{(1)}_{\text{M}}, \mathbf{m}^{(2)}_{\text{M}}, \ldots, \mathbf{m}^{(n_{\text{M}})}_{\text{M}})$ of the monitor posterior distribution $\rho_{\text{M}}(\mathbf{m}|\mathbf{d})$. These samples are then used to access the time-lapse statistical information. In particular, the correlation between baseline and monitor samples can be assessed by computing the covariance matrix:
\begin{equation}
   \Sigma_{\text{MB}} = \int_{\mathbb{M}}(\mathbf{m}_{\text{M}} - \mathbf{m}_{\text{M}}^{\text{mean}})^{\text{T}}(\mathbf{m}_{\text{B}} - \mathbf{m}_{\text{B}}^{\text{mean}})\rho_\mathbf{MB}(\mathbf{m}|\mathbf{d})\,d\mathbf{m}\,, 
\end{equation}
where $\rho_\mathbf{MB}(\mathbf{m}|\mathbf{d}) \equiv \rho(\mathbf{m}_{\text{M}}, \mathbf{m}_{\text{B}}| \mathbf{d}_{\text{M}}, \mathbf{d}_{\text{B}})$ is the associated joint probability distribution. In the remainder of the paper,  our focus will be on the comparison of the parallel and sequential Bayesian inversions using the HMC as the sampling method.

\subsection{Hamiltonian Monte Carlo}

The Hamiltonian Monte Carlo method incorporates the geometry of the target distribution to efficiently sample over high-dimensional model spaces~\citep{betancourt2018conceptual}.
In the HMC context, the model parameters $\mathbf{m}$ are interpreted as a set of particles following trajectories of a classical mechanical system. To complete the set of canonical variables, one defines the momenta $\mathbf{p} \in \mathbb{P}$ which is randomly sampled according to a Gaussian distribution with zero mean and a covariance given by the matrix mass $\mathbf{M}$. In this way, the  Hamiltonian system is subjected to a fake potential energy that mimics the negative value of the log-posterior distribution (Eq. \ref{post_dist}). Accordingly, the model space $\mathbb{M}$ is extended to a (fake) phase space $\mathbb{Z} = \mathbb{M}\times\mathbb{P}$, such that the posterior distribution is sampled over the canonical distribution 
\begin{equation}
    \rho(\mathbf{m}, \mathbf{p}) = \rho(\mathbf{p})\rho(\mathbf{m}|\mathbf{d}) \propto \exp{(-H(\mathbf{m}, \mathbf{p}))}\,,
\label{can_dist}
\end{equation}
with a Hamiltonian $H(\mathbf{m}, \mathbf{p})$ given by
\begin{equation}
    H(\mathbf{m}, \mathbf{p}) = \frac{1}{2}\mathbf{p}^\text{T}\mathbf{M}^{-1}\mathbf{p} - \log{\rho(\mathbf{m}|\mathbf{d})}\,.
\label{hamiltonian}
\end{equation}
To sample the distribution (Eq. \ref{can_dist}), one starts with a model $\mathbf{m}_0$ drawn from the prior distribution $\rho(\mathbf{m})$, which is then combined with initial random momenta $\mathbf{p}_0$ to generate the initial state of the system. Subsequently, the system evolves following the dynamics of Hamilton's equations~\citep{lemos_2018}: 
\begin{equation}
    \frac{d\mathbf{m}}{d\tau} = \frac{\mathbf{p}}{\mu}\,, \quad \frac{d\mathbf{p}}{d\tau} = \nabla \log{\rho(\mathbf{m}|\mathbf{d})}\,,
    \label{hamiltons_eq2}
\end{equation}
during a (artificial) time period $\tau$. The evolved state $(\mathbf{m}_\tau, \mathbf{p}_\tau)$ is accepted with a probability given by the Metropolis-Hasting criteria~\citep{metropolis1953}
\begin{equation}
    \mathrm{min}\left[1, \exp{(H(\mathbf{m}_0, \mathbf{p}_0)-H \left(\mathbf{m}_\tau, \mathbf{p}_\tau)\right)}\right]\,.
    \label{MH_criteria}
\end{equation}

Since we are interested in the estimated posterior (Eq. \ref{post_dist}), the canonical distribution is marginalized by discarding the momenta samples. Each sample $\mathbf{m}_\tau$ is stored and becomes the new initial position of the system, and it is again evolved according to the dynamical equations (Eq. \ref{hamiltons_eq2}) using new random momentum. This procedure is a single HMC step, which is then repeated $n_{\text{HMC}}$ times, producing a Markov chain with $n$ samples of the canonical distribution (Eq. \ref{can_dist}). We notice that only a fraction of the samples $(n \leq n_{\text{HMC}})$ is accepted, such that the acceptance rate $n/n_{\text{HMC}}$ is typically used as a control parameter to tune the HMC efficiency~\citep{leimkuhler94}. Therefore, it is recommended that the HMC parameters be adjusted to maintain high values ($n/n_{\text{HMC}} > 0.6$) of the acceptance rate. 

\subsubsection{Mass Matrix Tuning}

The mass matrix $\mathbb{M}$ has a strong impact on speeding up the phase space exploration and needs to be carefully adjusted~\citep {fichtner2021}. Improper choices can lead to high acceptance rates without an effective exploration of the phase space. On the other hand, assigning distinct mass values to different model parameters can be used to prioritize certain regions of the space~\citep{fichtner2018_2, Gebraad2020}, working similarly as regularizers in deterministic inversion~\citep{asnaashari2013, Asnaashari}. In particular, the optimal mass matrix choice for linear problems corresponds to the inverse of posterior covariance. However, the latter matrix is too large to be fully computed in practical problems and is unknown before the inversion. Thus, convenient choices are necessary to perform Bayesian inversion using HMC in high-dimensional problems.

We follow the strategy proposed in~\citep{DELIMA2023128618}, where each particle initially has a pseudo-mass $\mu$. After $k$ HMC steps, the mass matrix is updated to $\mathbf{M}^\prime = \boldsymbol{\Gamma}^{-1}\mathbf{M}$, where $\boldsymbol{\Gamma} = \text{diag}(\gamma_1, \gamma_2, \ldots, \gamma_{\text{dim}(\mathbb{M})})$ and
\begin{equation}
    \gamma_{i} = \gamma_{\text{min}} + (\gamma_{\text{max}} - \gamma_{\text{min}})\left(\frac{z - z_{\text{water}}}{z_{\text{max}} - z_{\text{water}}}\right)\,, \label{mass_adjusted}
\end{equation}
where $z_{\text{water}}$ and $z_{\text{max}}$ are the water layer depth and maximum model depth, respectively. The expression (Eq. \ref{mass_adjusted}) adjusts the mass of the particles according to its depth $z$ in the physical domain. This approach allows us to improve the HMC convergence using fewer forward calculations in comparison with the standard identity matrix choice. In addition, it also avoids additional Hessian approximation calculations~\citep{DELIMA2023128618}. In our study, we fix the initial mass to $\mu = 10$ and the minimum and maximum gamma values to $ \gamma_{\text{min}} = 1.0$ and $\gamma_{\text{max}} = 2.0$.

\section{Results}\label{sec:num}

\subsection{Experimental Setup}

\begin{figure}
    \centering
    \includegraphics[width=3.45in]{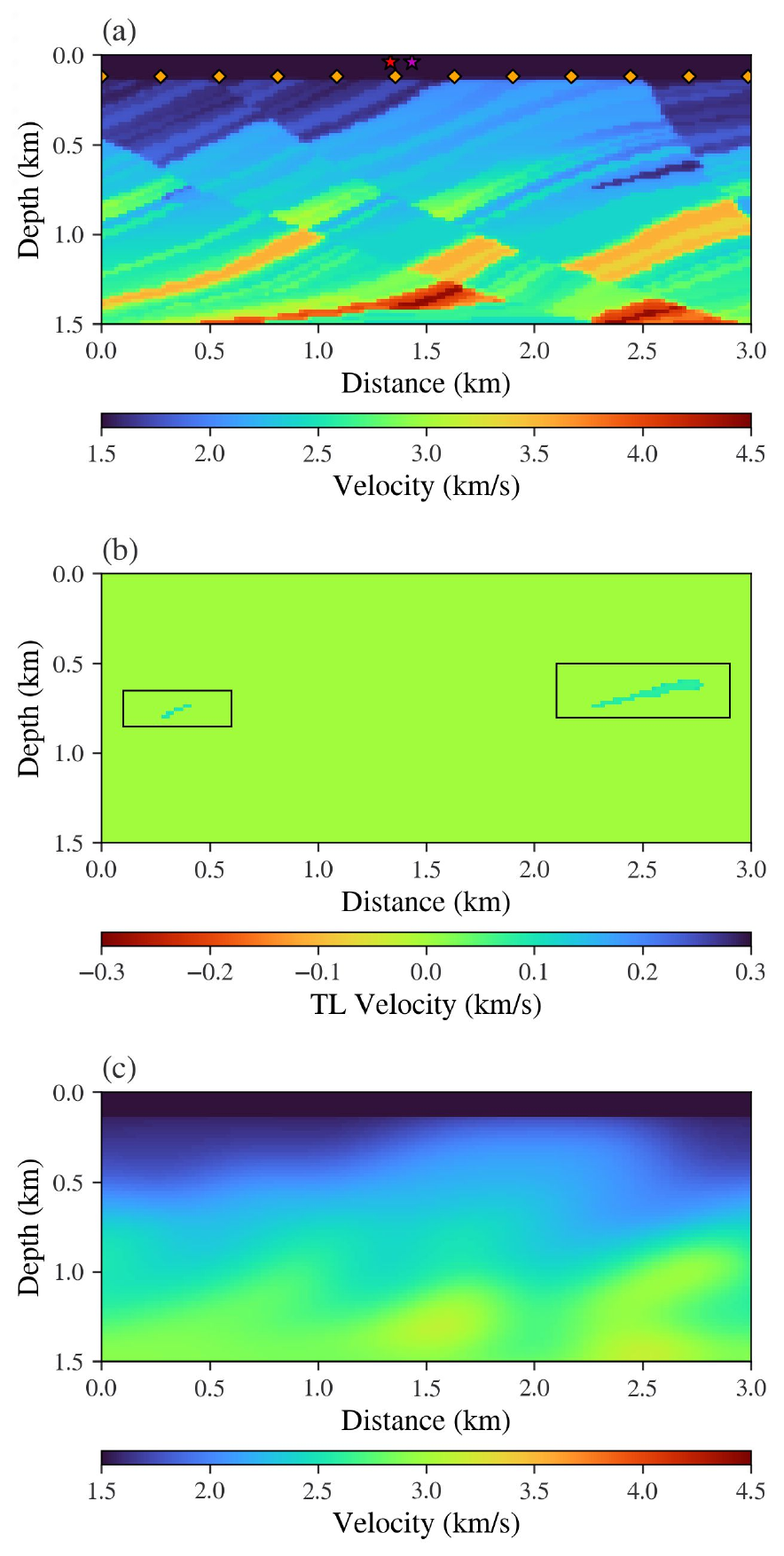}
    \caption{(a) The true baseline model with the red (magenta) star represents one of the ten sources used in the baseline (monitor perturbed) acquisition and the orange square represents some of the 200 receivers employed in our experiments. (b) The true time-lapse model. (c) The mean prior model used in the baseline estimation for both inversion strategies.}
    \label{fig:init_model}
\end{figure}

To compare both probabilistic time-lapse inversion approaches, our analysis is focused on the Marmousi model, which has been extensively used as a laboratory for time-lapse inversion studies~\citep{Asnaashari, LI2021103417, mardan2023, zhang3d2023}. The domain is a target zone (Figure \ref{fig:init_model}a) of the original model with 3.0 km of maximum offset and 1.5 km of depth, with a spatial resolution of 20 m in both directions. Therefore, the sampling model space has over $\sim 10^4$ dimensions for each survey.

We proceed as \citep{Asnaashari} and consider a time-lapse velocity model (Figure \ref{fig:init_model}b) which has $\sim 0.1$ km/s of velocity variation inside two gas reservoirs. These $\sim 5\%$ realistic changes represent the partial gas replacement by water. Initially, we simplify our analysis, considering a repeatable acquisition geometry (red star in Figure \ref{fig:init_model}a) between baseline and monitor surveys. After that, we consider a nonrepeatable scenario where the sources in the monitor survey (magenta star in Figure \ref{fig:init_model}a) are horizontally shifted by 100 m.

The synthetic baseline and monitor data were generated using 10 isotropic explosive sources (Ricker wavelet with 10 Hz of central frequency) at 40 m depth and recorded during 3.5 s by 200 receivers spaced every 20 m, deployed at 120 m depth. In accordance with Ref. \citep{curtis2021}, we also contaminate our seismic data with 0.1 standard deviation uncorrelated Gaussian noise. The corresponding forward modelling was performed by simulating a constant density acoustic wave propagation following a finite difference scheme with derivatives of eight-order in space and second-order in time. In addition, the gradient of the log-posterior (Eq. \ref{post_dist}) used in Hamilton's equations (Eq. \ref{hamiltons_eq2}) was computed using the adjoint state method~\citep{plessix2006, virieux2009}.

\subsection{Baseline Estimation and Monitor Prior Design}

\begin{figure*}
    \centering
    \includegraphics[width=\textwidth]{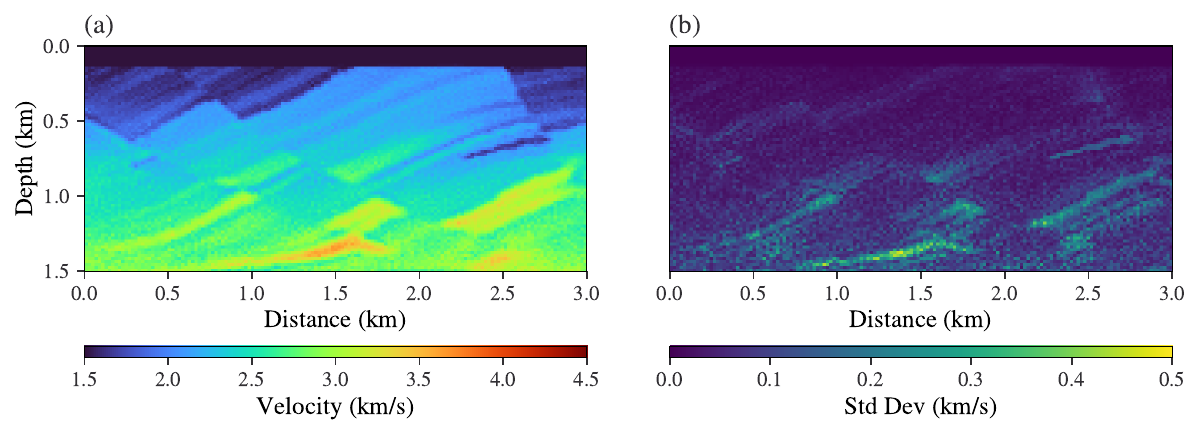}
    \caption{The mean (a) and the standard deviation (b) estimates of the baseline model using the HMC method.}
    \label{fig:bse_results}
\end{figure*}

The baseline posterior distribution $\rho_{\text{B}}(\mathbf{m}|\mathbf{d})$ was estimated from a prior probability $\rho_{\text{B}}(\mathbf{m})$ with mean model illustrated in Figure \ref{fig:init_model}c. This uninformative prior consists of independent and uniform distributions for each model parameter that assumes only the tomographic knowledge of the geological structure and extreme velocity values. About $800$ baseline samples were generated using 1200 HMC steps without discard, that is, no burn-in period was considered.

The Bayesian inversion results for the baseline survey are illustrated in Figure \ref{fig:bse_results}, where we focus on analysing the first statistical moment and the dispersion associated with respective posterior distribution. The main structures of the geological model are presented in the mean model (Figure \ref{fig:bse_results}a), particularly in the shallow part ($z < 1$ km), while only large-scale features were estimated at the deeper region ($z > 1$ km). As demonstrated by \citep{Gebraad2020} and \citep{DELIMA2023128618}, the noise data has a strong impact in assessing correct velocities mainly as the depth increases. Moreover, due to the limitations of the acquisition setup, it is difficult to detect dynamic structures around the bottom edge of the model. In fact, these features of the estimated baseline model are mapped as uncertainties in Figure \ref{fig:bse_results}b, where the standard deviation increases with depth and is higher in regions between discontinuities of the model, as in the region of reservoirs that we have particular interest. Similar uncertainty results using different sampling methods can be found in~\citep{izzatullah2021, curtis2020, curtis2021, DELIMA2023128618}.  

\begin{figure*}
    \centering
    \includegraphics[width=\textwidth]{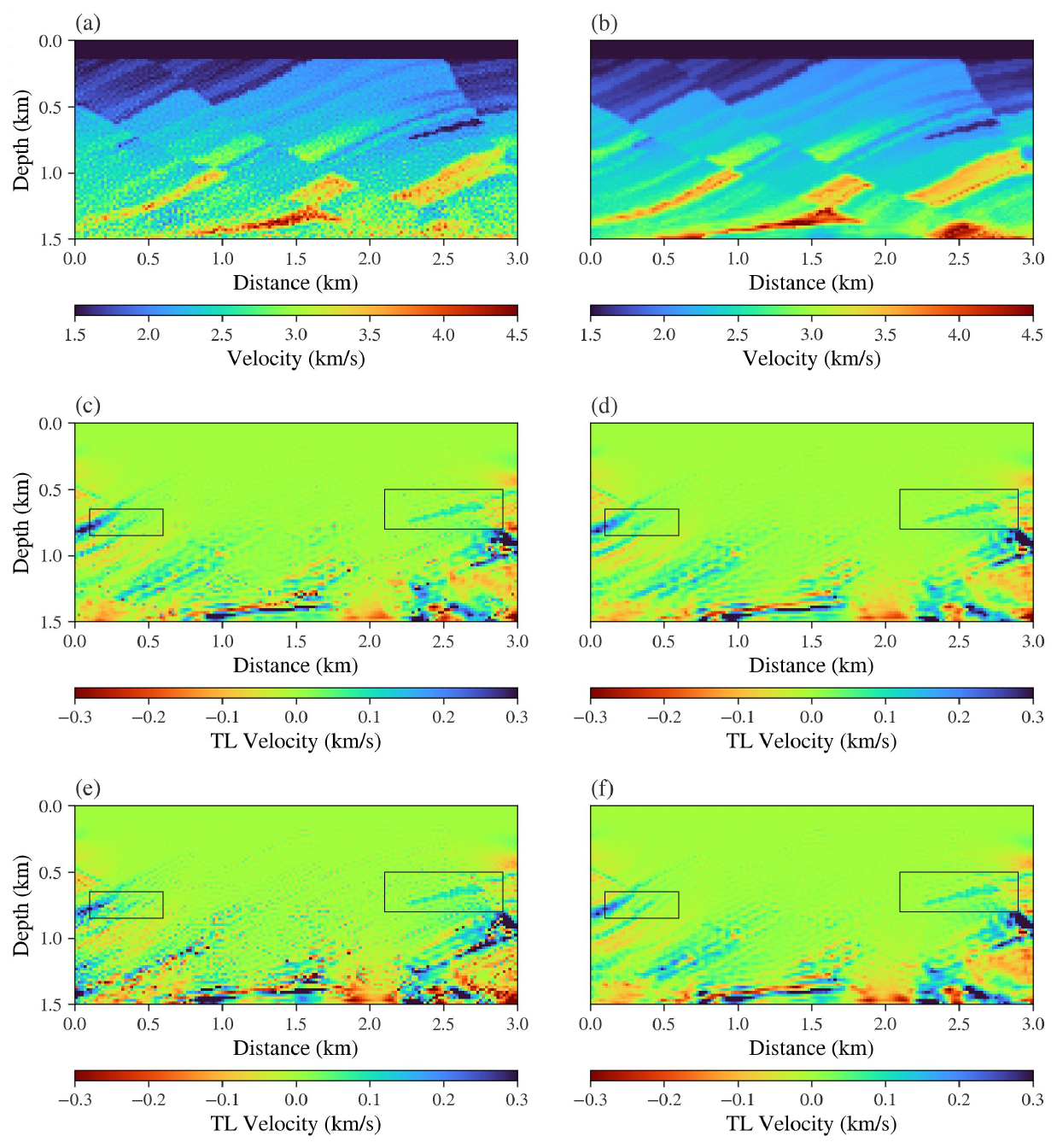}
    \caption{Comparison between deterministic and probabilistic inversions. The maximum a posteriori (MAP) baseline (\textbf{a}) and the model obtained using the deterministic inversion (\textbf{b}). Inverted time-lapse models using the deterministic version of the parallel (\textbf{left}) and sequential strategies (\textbf{right}) with perfect (\textbf{c, d}) and perturbed (\textbf{e, f}) acquisition geometry.}
    \label{fig:det_fwi}
\end{figure*}

As previously mentioned, deterministic methods fail to offer a precise framework for quantifying uncertainties of estimated velocity models. Nevertheless, we utilize them to compare the inversion results achieved through deterministic time-lapse workflows (parallel and sequential) with those obtained by employing their Bayesian probabilistic versions. For this purpose, we use the L-BFGS method \citep{lbfgs} to minimize the misfit function (argument of the likelihood function in Eq. \ref{likelihood}) in both cases, where the same experimental setup described at the beginning of this section was used. 
First, we compare the sample model that maximizes the posterior distribution $\rho_{\text{B}}(\mathbf{m}|\mathbf{d})$, that is, the maximum a posteriori (MAP) baseline model (Figure \ref{fig:det_fwi}a), with those obtained using the L-BFGS method (Figure \ref{fig:det_fwi}b) for the deterministic inversion. The two models (Figure \ref{fig:det_fwi}a and b) 
are quite similar mainly in the region of shorter offsets when $z < 1.2$ km, even though the L-BFGS uses more information (diagonal Hessian approximation) than the HMC to deliver smoother models.

For the parallel Bayesian inversion, the monitor posterior distribution $\rho_{\text{M}}(\mathbf{m}|\mathbf{d})$ was obtained from the same prior employed in the baseline estimation. However, the gain of information regarding the geological model due to baseline inversion can be useful to geometrically constrain the monitor estimation once the time-lapse changes are small and spatially localised. Therefore we use the baseline estimates to construct a prior distribution for the monitor survey in the sequential approach. Although the baseline posterior distribution $\rho_{\text{B}}(\mathbf{m}|\mathbf{d})$ possesses information concerning higher statistical moments, we define the monitor prior distribution $\rho_{\text{M}}(\mathbf{m})$ as a Gaussian approximation of the latter, that is, 
\begin{align}
    \rho_{\text{M}}(\mathbf{m}) &\equiv \rho_{\text{B}}(\mathbf{m}|\mathbf{d})\,, \nonumber \\ 
    &\approx \exp{\left[-\frac{1}{2}(\mathbf{m} - \mathbf{m}_{\text{B}}^{\text{mean}})^{\text{T}}\Sigma_{\text{B}}^{-1}(\mathbf{m} - \mathbf{m}_{\text{B}}^{\text{mean}})\right]}\,, 
\end{align}
where $\mathbf{m}_{\text{B}}^{\text{mean}}$ is the sample mean and $\Sigma_{\text{B}}$ is the sample baseline covariance matrix. In our calculations, only the diagonal elements of $\Sigma_{\text{B}}$ were considered. Additionally, we also use the prior baseline information to design a momentum distribution $\rho(\mathbf{p})$ where initial masses for model parameters located in the region around the reservoirs (black squares in Figure \ref{fig:init_model}b) were differently selected. We emphasize that our sequential approach differs from \citep{zhang2023bayesian} since the authors use the last few baseline samples as starting points to the monitor inversion instead of properly designing a new prior distribution.

\subsection{Repeatable Acquisition Geometry}

\begin{figure*}
    \centering
    \includegraphics[width=\textwidth]{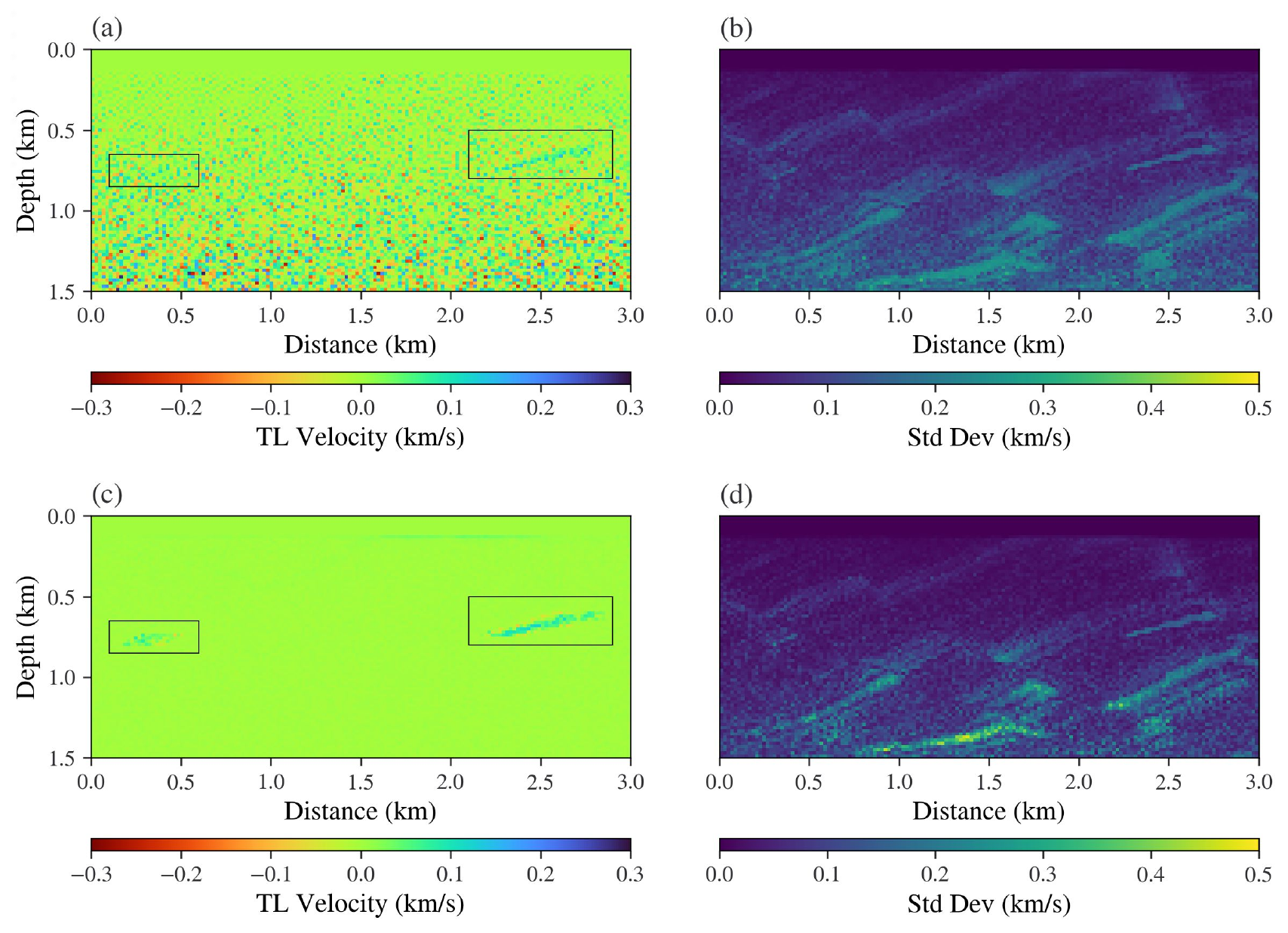}
    \caption{The mean (\textbf{left}) and the standard deviation (\textbf{right}) estimates of the time-lapse model with perfect acquisition geometry using the parallel (\textbf{a, b}) and sequential (\textbf{c, d}) Bayesian inversions.}
    \label{fig:tl_rep_results}
\end{figure*}

We compare the time-lapse inversion results using the parallel and sequential approaches in the perfect acquisition geometry scenario for both deterministic and probabilistic cases. The time-lapse posterior distribution was constructed using random pairs of baseline and monitor samples in both cases. Once again, we limit ourselves to investigating only the mean and standard deviation behaviour of the distribution, which are shown in Figure \ref{fig:tl_rep_results}. In the two strategies, the reservoirs are present in the time-lapse mean model, but the right reservoir is well-resolved in comparison with the left one. The small-scale structures (decoherent pixels with non-zero velocity values) that appear in the parallel scheme (Figure \ref{fig:tl_rep_results}a) can be associated with differences between baseline and monitor samples that do not cancel out on average. This kind of inversion artefact was recently reported by \citep{zhang2023bayesian} and also appears in the deterministic inversion (Figure \ref{fig:det_fwi}c) of previous studies~\citep{Asnaashari, zhou2021}. On the other hand, it is possible to identify clearly the velocity changes in the reservoir region when we employ the sequential strategy (Figure \ref{fig:tl_rep_results}c), even though it is difficult to recover the left-side reservoir in that case. These aspects of the sequential procedure also emerge in deterministic cases (Figure \ref{fig:det_fwi}d) and can be driven by affordable regularisation terms in the misfit function~\citep{Asnaashari}. 

\begin{figure*}
    \centering
    \includegraphics[width=\textwidth]{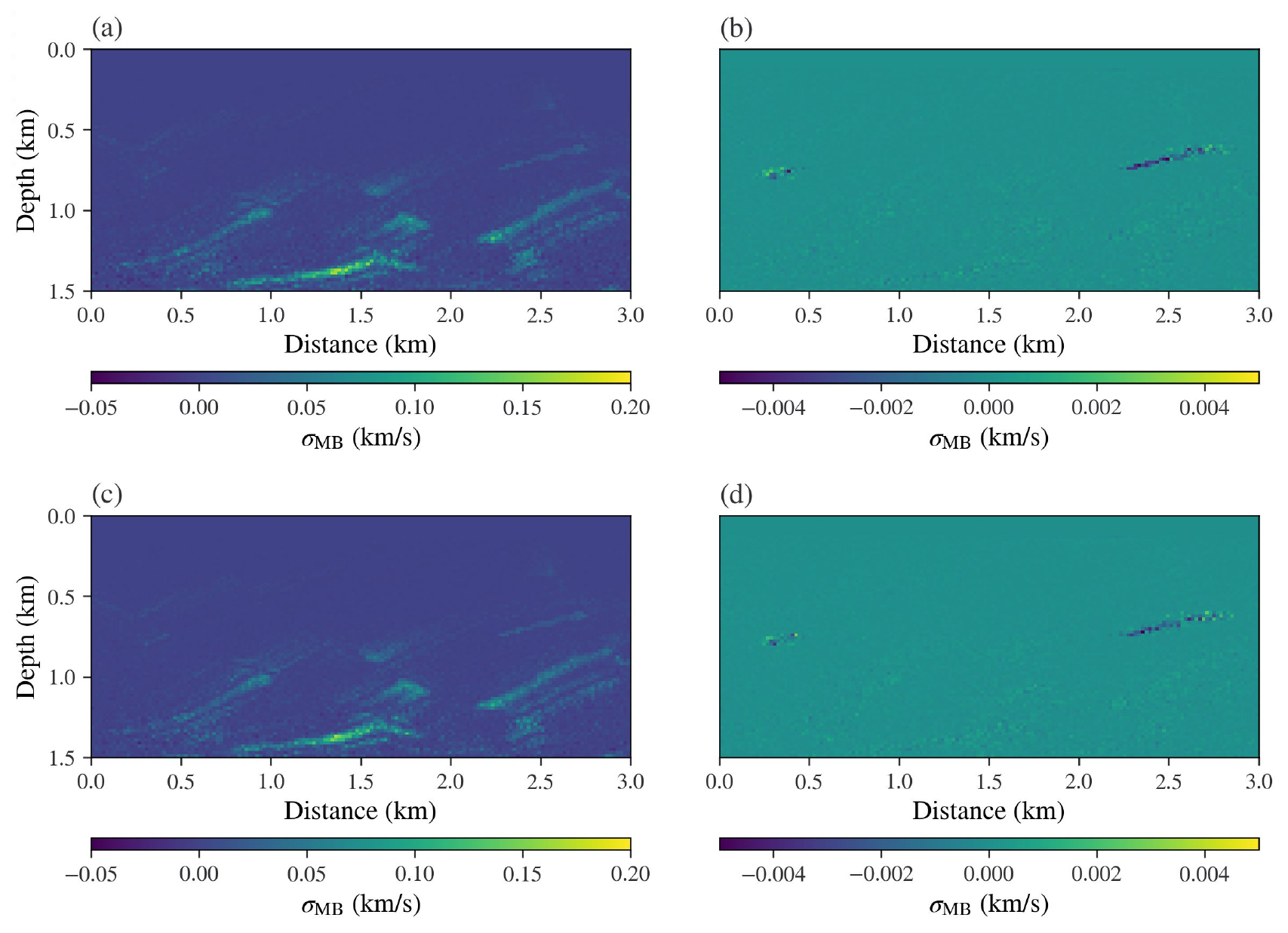}
    \caption{Estimated covariance between baseline and monitor samples in the parallel (\textbf{left}) and sequential (\textbf{right}) strategies with perfect (\textbf{a, b}) and perturbed (\textbf{c, d}) acquisition geometry.}
    \label{fig:correlation}
\end{figure*}

The mean estimates (Figure \ref{fig:tl_rep_results}a, c) are statistically meaningless without proper uncertainty quantification of results. The standard deviation $\sigma_{\text{TL}}$ of the time-lapse changes (Figures \ref{fig:tl_rep_results}b, d) was computed using the well-known first-order error propagation formula
\begin{equation}
    \sigma_{\text{TL}} = \sqrt{\sigma_{\text{M}}^2 + \sigma_{\text{B}}^2 - 2\sigma_{\text{MB}}}\,,\label{error_prop}
\end{equation}
where $\sigma_{\text{M}}$ and $\sigma_{\text{B}}$ are the standard deviation associated with the monitor and baseline inversion, respectively, and $\sigma_{\text{MB}}$ stands for the diagonal element of the covariance matrix between samples. In the light of Eq. \ref{error_prop}, errors in the baseline estimation will always be reflected as uncertainties in time-lapse images. Moreover, the propagated uncertainty also depends on the correlation degree between baseline and monitor samples. In this regard, uncertainties are lower in the parallel strategy, especially in the deeper part of the model due to the strong correlation with the baseline samples (Figure \ref{fig:correlation}a). This occurs because the sampling is performed from the same prior distribution. In the sequential strategy, monitor samples have lower uncertainty but weaker correlation (Figure \ref{fig:correlation}b). Therefore, the time-lapse images have similar uncertainties for both strategies.

\subsection{Nonrepeatable Acquisition Geometry}
\begin{figure*}
    \centering
    \includegraphics[width=\textwidth]{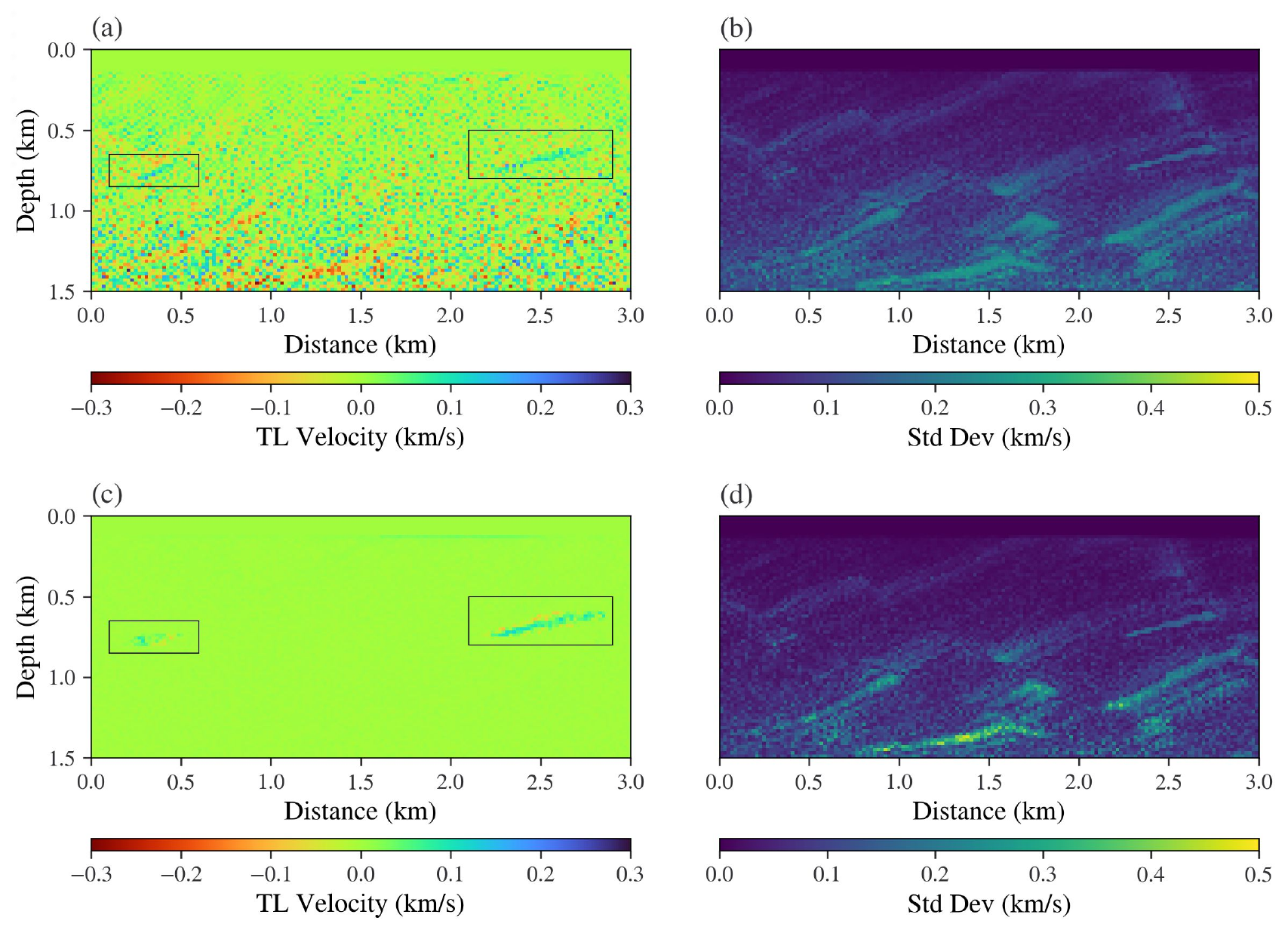}
    \caption{The mean (\textbf{left}) and the standard deviation (\textbf{right}) estimates of the time-lapse model with perturbed acquisition geometry using the parallel (\textbf{a, b}) and sequential (\textbf{c, d}) Bayesian inversions.}
    \label{fig:tl_nr_results}
\end{figure*}

We now consider the performance of the Bayesian inversion approaches when the acquisition geometry of the monitor survey differs from the baseline one. We first note that the correlation maps between monitor and baseline samples (Figure \ref{fig:correlation}c and d) are not sensitive to the nonrepeatable geometry regardless of the strategy used. This is evidence that the correlation between different surveys depends more on prior choice in model space $\mathbb{M}$ than slight changes in data space $\mathbb{D}$. Such similarities also appear in the standard deviation maps (Figure \ref{fig:tl_nr_results}c and d) for the parallel and sequential schemes, indicating that the estimated uncertainties to the time-lapse model are less affected by changes in source locations. However, the estimated mean time-lapse model in the parallel strategy (Figure \ref{fig:tl_nr_results}a) presents more artefacts when compared to the perfect geometry case (Figure \ref{fig:tl_rep_results}a), where the former has the presence of negative anomalies along the model that are paramount in the deeper part ($z < 1.0$ km). As it is well known in the deterministic framework~\citep{Asnaashari, zhou2021}, the sequential scheme and their extensions are more robust to nonrepeatability issues and as shown in Figure \ref{fig:tl_nr_results}c, the reservoirs can be better identified using this approach. 
\section{Discussion}

We have demonstrated in this work that exploring the posterior information of the baseline survey can benefit the estimation of well-resolved time-lapse changes. The occurrence of small-scale structures when the inversion is performed in parallel can be misunderstood as being geomechanical changes that blur the true velocity variation in the over and underburden regions. This behaviour for the time-lapse estimates (Figure \ref{fig:tl_rep_results}a, b) also were identified by \citep{zhang2023bayesian} using the sSVGD method. On the other hand, we argue that the knowledge regarding the baseline model can be used to design a suitable prior distribution in the monitor estimation, enabling better reservoir identifications. 

The strategies presented here depend on the quality and convergence of the baseline estimation. The correlation between samples (Figure \ref{fig:correlation}) shows that this dependence occurs differently in both methodologies, and it is crucial to decrease the uncertainties in the parallel when compared with the sequential inversion. Therefore, we observe a trade-off between the two approaches in terms of mean estimates and uncertainties. In fact, the results cannot be useful in practice when the uncertainties have the same magnitude order of the velocity changes. However, we also argue that the error in the baseline estimates can be alleviated by considering (1) a burn-in period that eliminates the initial chain model dependence and produces sample models with lower dispersion; (2) more samples that improve the convergences at the price of increasing the computational cost; (3) well-log information to better constrain vertical resolution the parameters in the reservoir's region or (4) data with reduced noise. Moreover, the sequential strategy is more robust to nonrepeatable errors, even though both inversion methods produce similar uncertainty estimations in perfect and perturbed acquisition geometry. Therefore, quantifying the uncertainty can help us discriminate inversion artefacts from true changes and make better decisions concerning reservoir management.


\section{Conclusion}
In this study, we propose a Bayesian sequential approach to time-lapse estimation using the Full Waveform Inversion (FWI). The idea involves integrating the baseline posterior information as prior knowledge in the monitor evaluation. Moreover, we use the Hamiltonian Monte Carlo (HMC) as a sampling method combined with an appropriate mass matrix design. We compare the proposed strategy with the parallel Bayesian inversion, where baseline and monitor estimations are conducted separately. These comparisons are performed with perfect and perturbed acquisition geometry, using a time-lapse Marmousi model as a study case. We concluded that the sequential strategy produces decorrelated samples between surveys and can provide more accurate time-lapse estimates with uncertainties of the same order as the parallel methodology.

\section*{Authors Contributions}

P. D. S. de Lima initiated the study, performed the simulations, analyzed the results, co-wrote, and reviewed the manuscript. M. S. Ferreira, G. Corso and J. M. de Araújo analyzed the results, co-wrote, and reviewed the manuscript.

\begin{acknowledgments}

The authors gratefully acknowledge support from Shell Brasil through the “New Computationally Scalable Methodologies for Target-Oriented 4D Seismic in Pre-Salt Reservoirs” project at Universidade Federal do Rio Grande do Norte (UFRN) and the strategic importance of the support given by ANP through the R\&D levy regulation. The authors thank the High-Performance Computing Center (NPAD) at UFRN for providing computational resources that made the present work possible. P. D. S. de Lima acknowledges the CAPES/PRINT program (process no. 88887.838338/2023-00). G. Corso acknowledges the CNPq Brazilian research agency for funding (grant 307907/2019-8). J. M. de Araújo acknowledges the CNPq Brazilian research agency for funding (grant 311589/2021-9).
\end{acknowledgments}

\section*{Data Availability Statement}

Data associated with this research are available and can be obtained by contacting the corresponding author.

\bibliographystyle{apsrev4-2}

\bibliography{bibliography}

\end{document}